# Emissions and energy efficiency on large-scale high performance computing facilities: ARCHER2 UK national supercomputing service case study

Adrian Jackson
EPCC, The University of Edinburgh
47 Potterrow, Edinburgh
EH8 9BT UK
a.jackson@epcc.ed.ac.uk

Alan Simpson
EPCC, The University of Edinburgh
47 Potterrow, Edinburgh
EH8 9BT UK
a.simpson@epcc.ed.ac.uk

Andrew Turner
EPCC, The University of Edinburgh
47 Potterrow, Edinburgh
EH8 9BT UK
a.turner@epcc.ed.ac.uk

## ABSTRACT

Large supercomputing facilities are critical to research in many areas that impact on decisions such as how to address the current climate emergency. For example, climate modelling, renewable energy facility design and new battery technologies. However, these systems themselves are a source of large amounts of emissions due to the embodied emissions associated with their construction, transport, and decommissioning; and the power consumption associated with running the facility. Recently, the UK National Supercomputing Service, ARCHER2, has been analysing the impact of the facility in terms of energy and emissions. Based on this work, we have made changes to the operation of the service that give a cumulative saving of more than 20% in power draw of the computational resources with all application benchmarks showing reduced power to solution. In this paper, we describe our analysis and the changes made to the operation of the service to improve its energy efficiency, and thereby reduce its climate impacts.

• Computing methodologies→Massively parallel and high-performance simulations.

## 1. INTRODUCTION

Large scale HPC systems have a key role to play in addressing the current climate emergency. They provide a digital laboratory where researchers can model and simulate areas of direct impact: climate modelling, renewable energy solutions, improved energy storage technologies, etc. while avoiding resource and emissions intensive physical experiments. However, these large systems are themselves large consumers of electricity and resources and, as such, are a source of significant emissions from their manufacture and installation as well as from their day-to-day operations [1]. Furthermore, the power draw of large HPC systems is significant and, particularly during times where there is competition for power on shared electricity grids; HPC systems must strive to be good "grid citizens". Finally, there are cost considerations. Historically, the cost of large scale HPC systems was dominated by the capital cost with the operational electricity costs a small component. This is no longer true, with lifetime electricity costs now matching or even exceeding the capital costs for large scale HPC systems in many countries.

In this paper we discuss the origin of emissions from a large scale HPC system, ARCHER2; describe the characteristics of the power draw of ARCHER2 broken down by different system components, and then review specific activities we have taken to improve the energy efficiency of the ARCHER2 system and their impact on application performance. We finish with some conclusions from the work and a description of future directions of interest.

### 1.1 ARCHER2

ARCHER2 [2] is the UK National Supercomputing Service funded by UK research councils (UKRI), managed by the Engineering and Physical Sciences Research Council (EPSRC) on behalf of UKRI with support provided by EPCC, hosting by the University of Edinburgh and the hardware provided by HPE.

The focus of this paper is the hardware component of the ARCHER2 service, which is a HPE Cray EX system with 750,080 compute cores. The system hardware is summarised in **Error! Reference source not found.**.

Table 1: ARCHER2 hardware summary

| | |
|---|---|
| 5860 compute nodes (750,080 compute cores) | 2× AMD EPYC[TM] 7842 2.25 GHz 64-core processors |
| | 256/512 GB DDR4 RAM |
| | 2 Slingshot 10 interconnect interfaces |
| Slingshot 10 interconnect | 768 Slingshot switches |
| | Dragonfly topology |
| Storage | 1PB NetApp storage |
| | 13.6 PB ClusterStor L300 (HDD-based) |
| | 1 PB ClusterStor E1000 (NVMe-based) |

ARCHER2 supports over 3000 users working on a broad range of research in the physical and environmental science areas with the major research areas being materials science, climate/ocean modelling, biomolecular modelling, engineering, mineral physics, seismology and plasma physics. It supports hundreds of different software packages looking at research problems that cannot be treated on other, smaller HPC facilities in the UK.

## 2. EMISSIONS

The hardware associated with the ARCHER2 service has two sources of climate emissions:

1. Operational emissions (scope 2 emissions) associated with the generation of electricity used to power the hardware and associated cooling/facility infrastructure.
2. Embodied emissions (scope 3 emissions) associated with the manufacture, shipping and decommission of the hardware.

There are no scope 1 emissions associated with the ARCHER2 hardware as there is no energy generation associated with the service – this is true for most large-scale HPC systems. A detailed audit of the emissions from ARCHER2 and emissions scenario modelling are underway and will be the subject of a future paper.

As a brief summary, in scenarios where the carbon density from scope 2 emissions is zero or very low (<30 g$CO_2$/kWh), the emissions from ARCHER2 are dominated by the scope 3, (embodied) emissions. In these cases, the best emissions efficiency is obtained by extracting the most output from each node hour (nodeh) for as long as possible. Anything that reduces the performance of applications on ARCHER2, will reduce the overall emissions efficiency. When the carbon intensity of scope 2 emissions is moderate (30-100 g$CO_2$/kWh), then the scope 2 and scope 3 emissions contribute roughly equally to the overall lifetime emissions. In this scenario, emissions efficiency is a combination of achieving energy efficiency and maximising application performance. However, if the emissions are dominated by scope 2 emissions (i.e. carbon intensity of the electricity used is high: >100 g$CO_2$/kWh), the emissions efficiency becomes dependent on the energy efficiency of the applications. In these scenarios, improving the operational energy efficiency by, for example, sacrificing application output per nodeh to improve application output per kWh, will improve the emissions efficiency. Evidentially, the approach to maximising emissions efficiency of large-scale HPC systems over their lifetime is dependent on the balance between the scope 2 and scope 3 emissions for the particular system: when scope 3 emissions dominate, optimise for application performance irrespective of energy efficiency; when scope 2 emissions dominate, optimise for energy efficiency, even if this has a detrimental impact on application performance. Many large-scale HPC systems will need to find a balance between application performance and energy efficiency to find a practical route to reduce emissions associated with operating such services.

## 3. ARCHER2 POWER DRAW

Irrespective of emissions, energy (or power) efficiency is still an important practical consideration for most large-scale HPC systems for many reasons, including:

- Limits on the amount of power that can be provided by the local power grid and competing demands for power. Data centres must be good grid citizens and be able to respond flexibly to fluctuating power demands, particularly during times of power shortages, where reducing the power draw of HPC systems can free up resources for other, critical, infrastructure.
- Desires to improve the cost efficiency of large scale HPC systems. Operational energy costs are now a major component of the lifetime costs of operating an HPC system.
- Higher power draw by HPC systems lead to higher cooling requirements increasing the overheads of running an HPC system.

Given the emissions profile for ARCHER2, and the additional practical reasons for improving energy efficiency, we have undertaken several initiatives to improve the energy efficiency of ARCHER2. This work was performed specifically within the context of reducing the power draw of ARCHER2 during Winter 2022/2023 when there were concerns about power shortages on the UK power grid.

To assess the impact of different initiatives on the power draw of the ARCHER2 service we need an understanding of the baseline power draw. To do this, we produced two sets of data:

1. Information on power draw of individual components.
2. Measurements of the baseline power draw over a few months.

### 3.1 Power draw of individual components

Table 2 shows the estimated idle and loaded per-component power draw. This data is a combination of measurements from the ARCHER2 system and estimates provided by the hardware vendor (HPE).

**Table 2: Estimated/measured power draw for different ARCHER2 system components. Italics indicates estimates.**

| Component | Notes | Idle (kW) [each] | Loaded (kW) [each] | Approx. % |
|---|---|---|---|---|
| Compute nodes | 5,860 nodes | 1,350 [0.23] | 3,000 [0.51] | 86% |
| Slingshot interconnect | 768 switches | 100-200 [0.10-0.25] | 200 [0.25] | 6% |
| Other Cabinet Overheads | 23 cabinets | 100-200 [4-9] | 200 [9] | 6% |
| Coolant Distribution Units | 6 CDUs | *96 [16]* | *96 [16]* | 3% |
| File systems | 5 file systems | *40 [8]* | *40 [8]* | 1% |
| Total | | 1,800 | 3,500 | |

Based on this information, we expect the power draw of the compute nodes to dominate on ARCHER2. Power draw associated with the interconnect switches are also a substantial component, but other system components (particularly storage) do not have a significant impact on the overall power draw of the system so can be discounted, at least initially, when considering ways to improve energy efficiency.

### 3.2 Baseline power draw measurements

We measured the baseline power draw of ARCHER2 compute cabinets (which includes all compute nodes and interconnect switches, approx. 90% of the total ARCHER2 power draw) from Dec 2021 to Apr 2022 – the timeline is shown in Figure 1. The mean power draw over this period was 3,220 kW. Compute node utilisation on ARCHER2 over all periods considered in this paper is consistently over 90% so the difference between idle node/switch power draw and loaded node/switch power draw does not need to be considered.

The mean value of 3,220 kW is lower than the sum of all loaded compute cabinet values from Table 2 (3,400 kW). This difference is partially due to the system not having 100% load (which is impossible to achieve due to scheduling overheads) and to differences in the power draw for different software running on the system. Basden and Turner [3] provides more details on the different power draw by different software applications.

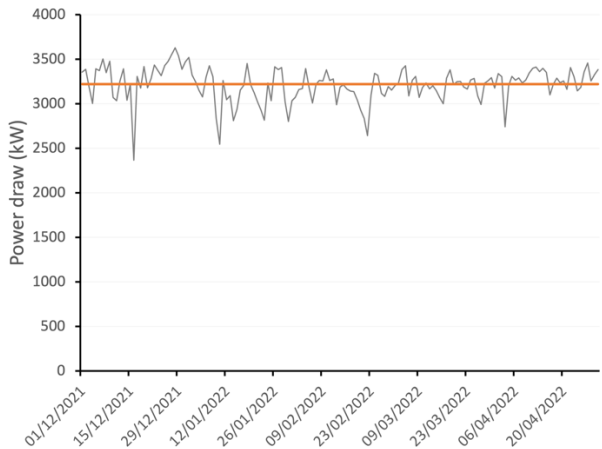

Figure 1: Measured power draw of ARCHER2 compute cabinets for Dec 2021 – Apr 2022. Orange line indicates mean power draw (3,220 kW).

## 4. IMPROVING ENERGY EFFICIENCY

Over the past 18 months we have investigated ways to improve the power usage of the ARCHER2 system, and thereby the energy efficiency, without requiring users of the service to take any action themselves. Ideally, we would reduce the power draw of the system without any impact on the performance of software, but this is typically not possible. In this paper we cover two strategies for improving the energy efficiency:

1. Change the compute node BIOS settings to move from Power Determinism mode to Performance Determinism mode.
2. Reduce the default CPU clock frequency from 2.25 GHz (with turbo-boost enabled) to 2.0 GHz (with no turbo-boost).

Both changes are relatively simple to make on a system-wide basis with no action required on individual users of the service. We summarise the impact of these changes on ARCHER2 power draw and on application performance in the remainder of this section.

### 4.1 Change BIOS to Performance Determinism mode

One setting available in the BIOS on compute nodes that use AMD CPUs is a choice between Power Determinism mode and Performance Determinism. A full description of the meaning and implication of these settings can be found in a technical report from AMD [4].

Figure 2 shows the impact of the change on the ARCHER2 compute cabinet power draw. The change was implemented across all compute nodes during May 2022 and led to a 7% reduction in the mean power draw of the ARCHER2 compute cabinet power draw, from 3,220 kW to 3,010 kW.

Table 3 reports the impact on performance and compute node energy consumption for several application benchmarks [5]. These show an impact of 1% or less on application performance and reductions in energy consumed on compute nodes for the applications between 6% and 10%.

### 4.2 Reduce CPU Clock Frequency to 2 GHz

The AMD CPUs on ARCHER2 allow the selection of different CPU frequencies, specifically 1.5 GHz, 2.0 GHz and 2.25 GHz. The highest frequency setting also enables the ability to turbo boost to higher frequencies. Reducing the CPU frequency reduces the rate at which the processor can execute instructions but, if application performance is limited by data transfer rates from memory to the processor rather than the rate of instruction execution, then this may not have a large detrimental effect on performance while reducing the power draw of the compute nodes. Many software applications that run on HPC systems such as ARCHER2 are memory bound in this way, rather than compute bound, so this could increase the energy efficiency of the system. We investigated the effect of reducing the CPU clock frequency to 2.0 GHz on performance and total energy use for a series of application benchmarks representing different research areas on ARCHER2, with results summarised in We also assessed the impact of this change in default CPU frequency on the power draw of the ARCHER2 compute cabinets. Figure shows the impact of the change on the ARCHER2 compute cabinet power draw. The change led to a reduction of the mean power draw of the ARCHER2 compute cabinet power draw from 3,010 kW to 2,530 kW, a total reduction of 21% compared to the original baseline power draw of 3,220 kW.

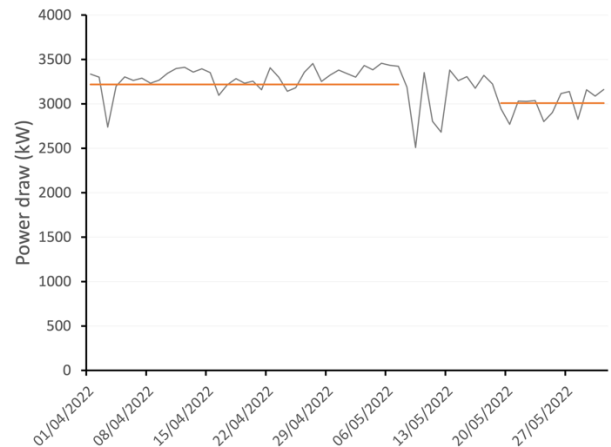

Figure 2: Measured power draw of ARCHER2 compute cabinets for Apr 2022 – May 2022. Orange line indicates mean power draw (3,220 kW before, 3,010 kW after).

Table 3: Performance and energy use comparison for application benchmarks with power determinism mode vs. performance determinism mode.

| Application benchmark | Nodes | Perf. ratio | Energy ratio |
|---|---|---|---|
| CASTEP Al Slab | 16 | 0.99 | 0.94 |
| OpenSBLI TGV $1024^3$ | 32 | 1.00 | 0.90 |
| VASP $TiO_2$ | 32 | 0.99 | 0.93 |

All the application benchmarks are more energy efficient at 2.0 GHz compared to 2.25 GHz, with energy savings ranging from 7% to 20%. Performance is more strongly affected than for the BIOS change described above, with reductions in performance of 5% to 26% when using the lower clock frequency. Some of these energy and performance reductions are larger than might be expected based on a change from 2.25 to 2.0 GHz. Further investigation revealed that most applications typically boost the CPU frequency to closer to 2.8 GHz in actual operation – explaining the larger range of the changes when limiting to 2.0 GHz. Based on this data, the decision was taken to improve the energy efficiency of the ARCHER2 service by setting the default CPU frequency to 2.0 GHz.

However, whilst the default was changed, users could revert these changes for their jobs. Furthermore, applications where the reduction in frequency is expected to have a large negative impact on performance (>10%) had their module setup altered to reset the CPU frequency to 2.25 GHz (with turbo boost enabled) automatically. Users were strongly encouraged to benchmark the effect of CPU frequency on their use of ARCHER2 and to choose an appropriate setting.

**Table 4: Performance and energy use comparison for application benchmarks with CPU frequency of 2.0 GHz compared to 2.25 GHz+turbo.**

| Application benchmark | Nodes | Perf. ratio | Energy ratio |
|---|---|---|---|
| CASTEP Al Slab[5] | 4 | 0.93 | 0.88 |
| CP2K $H_2O$ 2048 [6] | 4 | 0.91 | 0.93 |
| GROMACS 1400k [5] | 3 | 0.83 | 0.92 |
| LAMMPS Ethanol [7] | 4 | 0.74 | 0.92 |
| Nektar++ TGV 128 DoF [8] | 2 | 0.80 | 0.80 |
| ONETEP hBN-BP-hBN | 4 | 0.92 | 0.82 |
| VASP CdTe [5] | 8 | 0.95 | 0.88 |

We also assessed the impact of this change in default CPU frequency on the power draw of the ARCHER2 compute cabinets. Figure shows the impact of the change on the ARCHER2 compute cabinet power draw. The change led to a reduction of the mean power draw of the ARCHER2 compute cabinet power draw from 3,010 kW to 2,530 kW, a total reduction of 21% compared to the original baseline power draw of 3,220 kW.

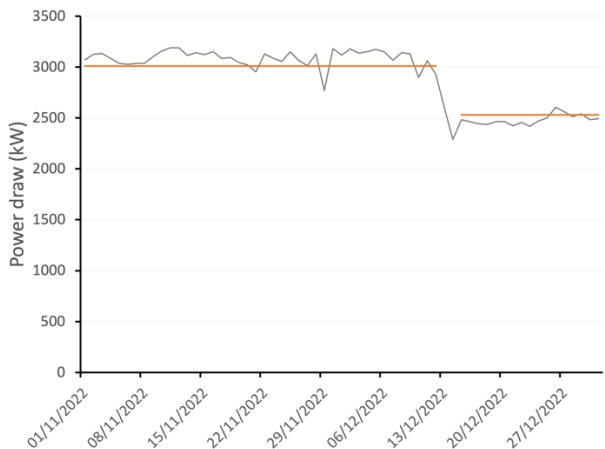

**Figure 3: Measured power draw of ARCHER2 compute cabinets for Nov 2022 – Dec 2022. Orange line indicates mean power draw (3,010 kW before, 2,530 kW after).**

## 5. CONCLUSIONS

We have made two low-overhead, system-wide changes to improve the energy efficiency of the ARCHER2 system which do not generally require user intervention or changes in behaviour, and which have a modest impact on application performance. Combined, these two changes reduce the energy draw of the ARCHER2 compute cabinets (which represent more than 90% of the power draw of the system) by an average of 690 kW, a reduction of 21% compared to the original average baseline power draw. This reduction in power draw freed up a substantial amount grid power capacity during a period of significant uncertainty in energy supplies in the UK and has resulted in significant savings in both scope 2 emissions and energy costs for the service. The change to the BIOS settings had a modest impact on power draw (210 kW, 6.5% reduction) and a negligible impact on the performance of application benchmarks (1% performance reduction). The default CPU frequency change had a much larger impact on power draw (480 kW, 15% reduction) and on application performance but can be reversed selectively by both the service operator and by users themselves on a per-application or per-job basis. All application benchmarks showed a reduction in total energy use at 2.0 GHz, 7% to 20% reduction, and the impact on performance varied from 5% to 26%.

To make correct choices about service operations in the areas discussed here, services must have a clear understanding of their priorities. For example, is the goal to maximise energy efficiency, to maximise emissions efficiency, to minimise running costs, to maximise application performance, or to achieve a balance between two or more different priorities? For ARCHER2, the primary goal was to maximise the energy efficiency due to potential power capacity shortages with a secondary goal of not having a large adverse impact on application performance.

During this work we noted that the power consumption of the most important components in terms of power draw is very high even when not being used for computational work. When compute nodes are not running user applications, they draw around 50% of power of a fully loaded compute node. The power draw of interconnect switches is steady at 200-250 W irrespective of system load. This means that to achieve good energy efficiency that utilisation of a system must be as close to 100% as possible and ideally over 90%.

Future papers will cover work we are undertaking to audit and model the emissions (scope 2 and scope 3) from ARCHER2 and large scale HPC systems more generally, looking at the impact on energy and emissions efficiency of replacing parts of modelling applications by AI-based approaches and investigating the impact of compiler and library choices on the energy efficiency of application benchmarks at different CPU frequencies.

## 6. ACKNOWLEDGMENTS

Our thanks to Martin Lafferty at HPE and Kieran Leach at EPCC, University of Edinburgh for making power monitoring from ARCHER2 cabinets and switches available for this work. This work used the ARCHER2 UK National Supercomputing Service (https://www.archer2.ac.uk). This research was supported by the NetZero Scoping project, which was funded by the UKRI Digital Research Programme on grant NERC (NE/W007134/1). AJ was supported by UK Research and Innovation under the EPSRC grant EP/T028351/1.